\PassOptionsToPackage{unicode}{hyperref}
\PassOptionsToPackage{hyphens}{url}
\PassOptionsToPackage{dvipsnames,svgnames,x11names}{xcolor}
\documentclass[
]{article}
\usepackage{xcolor}
\usepackage[top=10pc,bottom=10pc,left=11pc,right=11pc,heightrounded,top=1in,bottom=1in,left=1in,right=1in]{geometry}
\usepackage{amsmath,amssymb}
\usepackage{iftex}
\ifPDFTeX
  \usepackage[T1]{fontenc}
  \usepackage[utf8]{inputenc}
  \usepackage{textcomp} 
\else 
  \usepackage{unicode-math} 
  \defaultfontfeatures{Scale=MatchLowercase}
  \defaultfontfeatures[\rmfamily]{Ligatures=TeX,Scale=1}
\fi
\usepackage{lmodern}
\ifPDFTeX\else
\fi
\IfFileExists{upquote.sty}{\usepackage{upquote}}{}
\IfFileExists{microtype.sty}{
  \usepackage[]{microtype}
  \UseMicrotypeSet[protrusion]{basicmath} 
}{}
\usepackage{setspace}

\usepackage{longtable,booktabs,array}
\usepackage{calc} 
\usepackage{etoolbox}
\makeatletter
\patchcmd\longtable{\par}{\if@noskipsec\mbox{}\fi\par}{}{}
\makeatother
\IfFileExists{footnotehyper.sty}{\usepackage{footnotehyper}}{\usepackage{footnote}}
\makesavenoteenv{longtable}
\usepackage{graphicx}
\makeatletter
\newsavebox\pandoc@box
\newcommand*\pandocbounded[1]{
  \sbox\pandoc@box{#1}%
  \Gscale@div\@tempa{\textheight}{\dimexpr\ht\pandoc@box+\dp\pandoc@box\relax}%
  \Gscale@div\@tempb{\linewidth}{\wd\pandoc@box}%
  \ifdim\@tempb\p@<\@tempa\p@\let\@tempa\@tempb\fi
  \ifdim\@tempa\p@<\p@\scalebox{\@tempa}{\usebox\pandoc@box}%
  \else\usebox{\pandoc@box}%
  \fi%
}
\def\fps@figure{htbp}
\makeatother

\NewDocumentCommand\citeproctext{}{}
\NewDocumentCommand\citeproc{mm}{%
  \begingroup\def\citeproctext{#2}\cite{#1}\endgroup}
\makeatletter
 \let\@cite@ofmt\@firstofone
 \def\@biblabel#1{}
 \def\@cite#1#2{{#1\if@tempswa , #2\fi}}
\makeatother
\newlength{\cslhangindent}
\newlength{\csllabelwidth}
\newenvironment{CSLReferences}[2] 
 {\begin{list}{}{%
  \setlength{\itemindent}{0pt}
  \setlength{\leftmargin}{0pt}
  \setlength{\parsep}{0pt}
  \ifodd #1
   \setlength{\leftmargin}{\cslhangindent}
   \setlength{\itemindent}{-1\cslhangindent}
  \fi
  \setlength{\itemsep}{#2\baselineskip}}}
 {\end{list}}
\usepackage{calc}

\providecommand{\tightlist}{%
  \setlength{\itemsep}{0pt}\setlength{\parskip}{0pt}}

\usepackage[all,defaultlines=2]{nowidow}

\usepackage{enumitem}
\setlist[1]{labelindent=\parindent}
\setlist[itemize]{leftmargin=*}
\setlist[enumerate]{leftmargin=*}

\setlist[description]{style=unboxed}

\usepackage[titles]{tocloft}

\AtBeginEnvironment{longtable}{\setlist[itemize]{nosep, wide=0pt, leftmargin=*, before=\vspace*{-\baselineskip}, after=\vspace*{-\baselineskip}}}

\usepackage{setspace}

\usepackage[overload]{textcase}
\usepackage[runin]{abstract}

\abslabeldelim{\quad}
\usepackage{titling}
\pretitle{\par\vskip 5em \begin{flushleft}\LARGE\sffamily\bfseries}
\posttitle{\par\end{flushleft}\vskip 0.75em}

\renewcommand{\and}{\end{tabular} \hskip 3em \begin{tabular}[t]{@{\hspace{0em}}l@{}}}
\preauthor{\begin{flushleft}
           \lineskip 1.5em 
           \begin{tabular}[t]{@{\hspace{0em}}l@{}}}
\postauthor{\end{tabular}\par\end{flushleft}}

\predate{}
\postdate{}

\newcommand{\published}[1]{%
   \gdef\puB{#1}}
   \newcommand{\puB}{}

\usepackage{titlesec}
\titleformat*{\section}{\Large\sffamily\bfseries\raggedright}
\titleformat*{\subsection}{\large\sffamily\bfseries\raggedright}
\titleformat*{\subsubsection}{\normalsize\sffamily\bfseries\raggedright}
\titleformat*{\paragraph}{\small\sffamily\bfseries\raggedright}

\titlespacing*{\section}{0em}{2em}{0.1em}
\titlespacing*{\subsection}{0em}{1.25em}{0.1em}
\titlespacing*{\subsubsection}{0em}{0.75em}{0em}

\usepackage[bottom, flushmargin]{footmisc}

\usepackage[font={small,sf}, labelfont={small,sf,bf}]{caption}

\usepackage{pdflscape}
\newcommand{\blandscape}{\begin{landscape}}
\newcommand{\elandscape}{\end{landscape}}

\usepackage{bbm}
\let\origmathbb\mathbb
\renewcommand{\mathbb}[1]{\ifnum\pdfstrcmp{#1}{1}=0 \mathbbm{1}\else\origmathbb{#1}\fi}
\usepackage{xcolor}
\definecolor{coloraccent}{HTML}{107895}
\usepackage[section]{placeins}
\usepackage{float}
\usepackage{tabularray}
\usepackage[normalem]{ulem}
\usepackage{graphicx}
\usepackage{rotating}
\UseTblrLibrary{siunitx}
\NewTableCommand{\tinytableDefineColor}[3]{\definecolor{#1}{#2}{#3}}

\makeatletter
\@ifpackageloaded{caption}{}{\usepackage{caption}}
\AtBeginDocument{%
\ifdefined\contentsname
  \renewcommand*\contentsname{Table of contents}
\else
  \newcommand\contentsname{Table of contents}
\fi
\ifdefined\listfigurename
  \renewcommand*\listfigurename{List of Figures}
\else
  \newcommand\listfigurename{List of Figures}
\fi
\ifdefined\listtablename
  \renewcommand*\listtablename{List of Tables}
\else
  \newcommand\listtablename{List of Tables}
\fi
\ifdefined\figurename
  \renewcommand*\figurename{Figure}
\else
  \newcommand\figurename{Figure}
\fi
\ifdefined\tablename
  \renewcommand*\tablename{Table}
\else
  \newcommand\tablename{Table}
\fi
}
\@ifpackageloaded{float}{}{\usepackage{float}}
\floatstyle{ruled}
\@ifundefined{c@chapter}{\newfloat{codelisting}{h}{lop}}{\newfloat{codelisting}{h}{lop}[chapter]}
\floatname{codelisting}{Listing}

\usepackage{amsthm}
\theoremstyle{plain}
\newtheorem{algorithm}{Algorithm}[section]
\theoremstyle{remark}
\AtBeginDocument{}

\makeatother
\makeatletter
\@ifpackageloaded{caption}{}{\usepackage{caption}}
\@ifpackageloaded{subcaption}{}{\usepackage{subcaption}}
\makeatother
\usepackage{bookmark}
\IfFileExists{xurl.sty}{\usepackage{xurl}}{} 
\hypersetup{
  pdftitle={Marriage and Divorce in Continuous Time},
  pdfauthor={Kazuharu Yanagimoto},
  colorlinks=true,
  linkcolor={coloraccent},
  filecolor={Maroon},
  citecolor={coloraccent},
  urlcolor={coloraccent},
  pdfcreator={LaTeX via pandoc}}

\usepackage{orcidlink}  

\title{Marriage and Divorce in Continuous Time\thanks{This work was
supported by JSPS KAKENHI Grant Number 25K23111.}}

\author{
{\large Kazuharu Yanagimoto~\orcidlink{0009-0007-1967-8304}}%
 \\%
Kobe University \\%
{\footnotesize \url{yanagimoto@econ.kobe-u.ac.jp}} \and
}

\date{}

\usepackage{url}
\begin{document}
\published{\textbf{Monday, February 23, 2026} \\ {\scriptsize Access the
code, data, and analysis at
\url{https://github.com/kazuyanagimoto/spruce-shark}}}

\maketitle

\begin{abstract}
This paper reformulates the Greenwood and Guner
(\citeproc{ref-greenwood2009}{2009}) marriage and divorce model in
continuous time using the HACT methods of Achdou et al.
(\citeproc{ref-achdou2022}{2022}). Replacing the AR(1) match quality
process with an Ornstein-Uhlenbeck process yields a tridiagonal
generator, reducing the computational complexity of both the value
function and stationary distribution calculations from quadratic to
linear in the number of grid points. The continuous-time model closely
replicates the discrete-time equilibrium across all key outcomes,
including the share of married households, the marriage rate, and the
divorce rate, while achieving substantial gains in computation time and
memory usage.
\end{abstract}
\vskip 3em


\setstretch{1.25}
\vskip -3em \hspace{\parindent}{\sffamily\footnotesize\bfseries\MakeUppercase{JEL Codes}}\quad {\sffamily\small C63; D13; J12}\vskip 3em
\vskip -3em \hspace{\parindent}{\sffamily\footnotesize\bfseries\MakeUppercase{Keywords}}\quad {\sffamily\small Marriage; Divorce; Continuous Time; Search and Matching}\vskip 3em

\section{Introduction}\label{introduction}

Quantitative models of marriage and divorce have become a central tool
for understanding long-run demographic change and evaluating family
policy (e.g., \citeproc{ref-greenwood2009}{Greenwood and Guner 2009};
\citeproc{ref-greenwood2016}{Greenwood et al. 2016};
\citeproc{ref-voena2015}{Voena 2015}; \citeproc{ref-reynoso2024}{Reynoso
2024}). In the canonical framework of Greenwood and Guner
(\citeproc{ref-greenwood2009}{2009}), the postwar decline in marriage
rates and rise in divorce rates are attributed to two secular forces:
rising wages, which raise the opportunity cost of home production, and
falling prices of home goods, which reduce the gains from specialization
within marriage. By embedding a search and matching structure with
stochastic match quality into a household allocation model, Greenwood
and Guner (\citeproc{ref-greenwood2009}{2009}) provide a unified
quantitative account of U.S. marital dynamics from 1950 to 2000.

Their model, like most quantitative family economics models, is
formulated in discrete time. The match quality of married couples
follows an AR(1) process discretized via the method of Tauchen
(\citeproc{ref-tauchen1986}{1986}), and the stationary distribution is
computed by iterating on a full transition matrix. This approach is
straightforward to implement but scales poorly: the Tauchen
approximation produces a dense \(N \times N\) transition matrix, so that
both value function iteration and the stationary distribution
calculation require \(O(N^2)\) operations in the number of grid points
\(N\).

Recent advances in continuous-time methods for heterogeneous agent
models (e.g., \citeproc{ref-achdou2022}{Achdou et al. 2022};
\citeproc{ref-kaplan2018}{Kaplan et al. 2018}) suggest an alternative.
When the match quality follows an Ornstein--Uhlenbeck process, the
continuous-time counterpart of the AR(1) process, the infinitesimal
generator is a tridiagonal matrix, and both the Hamilton-Jacobi-Bellman
(HJB) equation and the Kolmogorov Forward equation can be solved in
\(O(N)\) time. This reduction from quadratic to linear complexity can
yield substantial computational gains, particularly when fine grids are
needed to resolve the divorce threshold accurately.

This paper reformulates the Greenwood and Guner
(\citeproc{ref-greenwood2009}{2009}) marriage and divorce model in
continuous time and demonstrates that the continuous-time method
replicates the discrete-time equilibrium while offering significant
computational advantages. The continuous-time formulation replaces the
AR(1) match quality process with an OU process, the Bellman equations
with HJB equations incorporating a variational inequality for endogenous
divorce (cf. \citeproc{ref-mellior2024}{Mellior and Shibayama 2024}),
and the stationary distribution iteration with a Kolmogorov Forward
equation. By applying the HACT (Heterogeneous Agent Continuous Time)
approach of Achdou et al. (\citeproc{ref-achdou2022}{2022}), all
numerical operations reduce to tridiagonal linear systems.

The quantitative results show that the continuous-time model closely
tracks the discrete-time equilibrium across all key outcomes: the share
of married households, the marriage rate, and the divorce rate.
Benchmark comparisons confirm that computation time and memory scale as
\(O(N)\) for the continuous-time method and \(O(N^2)\) for the
discrete-time method, consistent with the theoretical complexity orders.

The paper is organized as follows. Section~\ref{sec-model} presents the
discrete-time and continuous-time marriage models.
Section~\ref{sec-estimation} describes the calibration strategy and
reports the simulation results and computational benchmarks.
Section~\ref{sec-conclusion} concludes. Section~\ref{sec-apdx-comp}
provides computational details on the numerical algorithms.

\section{Model}\label{sec-model}

To develop a continuous-time search and matching model of marriage, I
build on the canonical framework of Greenwood and Guner
(\citeproc{ref-greenwood2009}{2009}). I begin by presenting their
discrete-time formulation and then extend it to continuous time.

\subsection{Discrete-time Model}\label{discrete-time-model}

The economy is populated by a continuum of infinitely-lived households.
Each household has a marital status, single or married. In each period,
each single meets a potential partner with a match quality
\(b \in \mathcal{B}\).\footnote{This match quality \(b\) is also called
  ``bliss shock'' in Greenwood and Guner
  (\citeproc{ref-greenwood2009}{2009}).} They marry if the expected
value of being married exceeds that of being single. Each married
household receives a flow utility \(b\) from the match quality, and can
divorce at any time. The match quality evolves stochastically over time,
and households make their decisions based on the current match quality
and the expected future values. At the end of each period, households
exit the economy with probability \(\delta\), and new households (whose
mass is \(\delta\)) enter the economy as singles.

\subsubsection*{Household Allocation}\label{household-allocation}
\addcontentsline{toc}{subsubsection}{Household Allocation}

Let \(z \in \{1, 2\}\) denote the number of adults in the household,
i.e., \(z = 1\) for singles and \(z = 2\) for married household. Each
household is endowed with \(z\) units of time, which can be allocated to
market work \(l\) and home hours \(h\). Households obtain utility from
consumption \(c\) and home production \(n\), which is a composite good
from home goods \(d\) and home hours \(h\). The household's problem is
given by

\begin{equation}\protect\phantomsection\label{eq-utility-dt}{
\max_{c, d, h} \alpha \log \left(\frac{c - \overline{c}}{z^\phi}\right) + \frac{1-\alpha}{\zeta} \left(\frac{n}{z^{\phi}}\right)^{\zeta}
}\end{equation} subject to

\[
\begin{aligned}
c + wpd &= w(z-h) \\
n & = \left(\theta d^{\kappa} + (1-\theta)h^{\kappa}\right)^{\frac{1}{\kappa}},
\end{aligned}
\] where \(w\) is the wage rate and \(p\) is the price of home goods
(relative to the wage). The parameter \(\overline{c}\) is the minimum
consumption level, and \(\phi\) captures the scale effect of household
size on utility. The parameter \(\kappa\) captures the substitutability
between home goods and home hours in home production, and \(\theta\) is
the share of home goods in home production. The indirect utility
function of Eq.~\ref{eq-utility-dt} is denoted by \(v(p, w, z)\).

\subsubsection*{Match Quality}\label{match-quality}
\addcontentsline{toc}{subsubsection}{Match Quality}

Single households meet potential partners with match quality \(b\) drawn
from a normal distribution:

\begin{equation}\protect\phantomsection\label{eq-dist-match-single}{
b \sim \mathcal{N}(\mu_s, \sigma_s^2),
}\end{equation} where \(\mu_s\) and \(\sigma_s^2\) are the mean and
variance of the match quality of singles. The cumulative distribution
function of the match quality of singles is denoted by \(F(b)\). The
match quality of married households evolves according to an AR(1)
process:

\begin{equation}\protect\phantomsection\label{eq-ar1}{
b = (1-\varrho)\mu_m + \varrho b_{-1} + \sigma_{m}\sqrt{1-\varrho^2}\xi, \quad \xi \sim \mathcal{N}(0, 1).
}\end{equation} The cumulative distribution function of the match
quality of married households is denoted by \(G(b'|b)\).

\subsubsection*{Value Functions}\label{value-functions}
\addcontentsline{toc}{subsubsection}{Value Functions}

The value functions of married and single households, denoted by
\(V(b)\) and \(W\), are given by

\begin{equation}\protect\phantomsection\label{eq-bellman-dt}{
\begin{aligned}
V(b) &= v(p, w, 2) + b + \beta \int_{\mathcal{B}} \max\{V(b'), W\} \,dG(b'|b), \\
W &= v(p, w, 1) + \beta \int_{\mathcal{B}} \max\{V(b), W\} \,dF(b).
\end{aligned}
}\end{equation}

The effective discount rate \(\beta := \tilde{\beta} (1 - \delta)\) is
the product of the discount factor \(\tilde{\beta}\) and the survival
probability \(1 - \delta\). The household chooses to marry if
\(V(b) \geq W\) and remains single otherwise. Married households choose
to divorce if \(V(b) < W\) and remain married otherwise.

\subsubsection*{Stationary Distribution}\label{stationary-distribution}
\addcontentsline{toc}{subsubsection}{Stationary Distribution}

Let \(s_{-1}\) and \(M_{-1}(b)\) be the density of singles and married
households with match quality \(b\) in the previous period. Then, the
density of singles and married households with match quality \(b\) in
the current period, denoted by \(s\) and \(M(b)\), are given by

\begin{equation}\protect\phantomsection\label{eq-stationary-dt}{
\begin{aligned}
s &= \delta + (1-\delta)s_{-1}\int_{\mathcal{B}} \mathbb{1}\left\{V(b) < W\right\} \,dF(b)\\
& + (1-\delta)(1-s_{-1})\int_{\mathcal{B}}\int_{\mathcal{B}}^{b} \mathbb{1}\left\{V(b') < W\right\} \,dG(b'|b_{-1})\,dM_{-1}(b_{-1}),\\
M(b) &= (1-\delta)s_{-1}\int_{\mathcal{B}}^{b} \mathbb{1}\left\{V(b') \geq W\right\} \,dF(b')\\
& + (1-\delta)(1-s_{-1})\int_{\mathcal{B}}\int_{\mathcal{B}}^{b}\mathbb{1}\left\{V(b') \geq W\right\} \,dG(b'|b_{-1})\,dM_{-1}(b_{-1}).
\end{aligned}
}\end{equation}

The stationary distribution is the fixed point of
Eq.~\ref{eq-stationary-dt}, i.e., \(s = s_{-1}\) and
\(M(b) = M_{-1}(b)\) for all \(b \in \mathcal{B}\).

\subsubsection*{Equilibrium}\label{equilibrium}
\addcontentsline{toc}{subsubsection}{Equilibrium}

To close the model, the production side is defined as a linear
technology, \(y = wl\). The price of home goods \(p\) is exogenously
given. A stationary equilibrium is a tuple \(\{V(b), W, s, M(b)\}\) such
that

\begin{enumerate}
\def\labelenumi{\arabic{enumi}.}
\tightlist
\item
  \(V(b)\) and \(W\) satisfy Eq.~\ref{eq-bellman-dt}
\item
  \(s\) and \(M(b)\) constitute a stationary distribution satisfying
  Eq.~\ref{eq-stationary-dt} with \(s = s_{-1}\) and
  \(M(b) = M_{-1}(b)\) for all \(b \in \mathcal{B}\).
\end{enumerate}

\subsection{Continuous-time Model}\label{continuous-time-model}

Consider a continuous-time version of the model. While the household
allocation is the same as in the discrete-time model, the timing of
events is different. In continuous time, households exit the economy
with a constant hazard rate \(\nu\), and new households enter the
economy as singles with the same hazard rate \(\nu\). Each single
household meets potential partners with a Poisson arrival rate
\(\lambda\), and the match quality is drawn from the same distribution
as in the discrete-time model. The match quality of married households
evolves according to an Ornstein--Uhlenbeck (OU) process. The value
functions satisfy Hamilton-Jacobi-Bellman (HJB) equations, and the
stationary distribution is the solution to the Kolmogorov Forward (KF)
equation.

\subsubsection*{Match Quality}\label{match-quality-1}
\addcontentsline{toc}{subsubsection}{Match Quality}

Single households meet potential partners with Poisson arrival rate
\(\lambda\), and the match quality \(b_t\) is drawn from a normal
distribution Eq.~\ref{eq-dist-match-single}, whose CDF and PDF are
\(F(b)\) and \(f(b)\). The match quality of married households evolves
according to an OU process:

\begin{equation}\protect\phantomsection\label{eq-ou-process}{
d b_t = \eta(\mu_m - b_t) dt + \sigma_{m}\sqrt{2\eta}\, dB_t,
}\end{equation} where \(B_t\) is a standard Brownian motion. Under this
parametrization, the stationary distribution of \(b_t\) is
\(\mathcal{N}(\mu_m, \sigma_m^2)\), so \(\mu_m\) and \(\sigma_m\) have
the same interpretation as in the discrete-time AR(1) specification
Eq.~\ref{eq-ar1}.

\subsubsection*{Value Functions}\label{value-functions-1}
\addcontentsline{toc}{subsubsection}{Value Functions}

Let \(V(b)\) and \(W\) be the value functions of married and single
households. Single households choose to marry if \(V(b) \geq W\) and
remain single otherwise. Married households choose to divorce if
\(V(b) < W\) and remain married otherwise. This implies the following
variational inequality:

\[
V(b) \geq W \quad \text{for all }b \in \mathcal{B},
\] which defines an endogenous marriage/divorce threshold:

\[
b^* = \inf\{b \in \mathcal{B} : V(b) \geq W\}.
\]

The value functions satisfy the following HJB equations:

\begin{equation}\protect\phantomsection\label{eq-hjb-single}{
\rho W = v(p, w, 1) + \lambda \int_{b^*}^{\infty} V(b') - W \,dF(b'),
}\end{equation}

\begin{equation}\protect\phantomsection\label{eq-hjb-married}{
\rho V(b) = v(p, w, 2) + b + \eta(\mu_m - b) \frac{\partial V(b)}{\partial b} + \eta\sigma_m^2 \frac{\partial^2 V(b)}{\partial b^2}.
}\end{equation}

The effective discount rate \(\rho\) is the sum of the discount rate and
the exit rate, \(\rho = \tilde{\rho} + \nu\).

\subsubsection*{Stationary
Distribution}\label{stationary-distribution-1}
\addcontentsline{toc}{subsubsection}{Stationary Distribution}

The stationary distribution is the solution to the Kolmogorov Forward
equation, which is given by

\begin{equation}\protect\phantomsection\label{eq-kf}{
0 = -\frac{\partial}{\partial b}\left[\eta(\mu_m - b)M(b)\right] + \eta\sigma_m^2 \frac{\partial^2 M(b)}{\partial b^2} + \lambda s f(b) - \nu M(b) \quad \text{for } b > b^*,
}\end{equation} where \(M(b)\) is the density of married households with
match quality \(b\), and \(s\) is the mass of single households,

\begin{equation}\protect\phantomsection\label{eq-population-ct}{
s = 1 - \int_{b^*}^{\infty} M(b) \, db.
}\end{equation}

\subsubsection*{Equilibrium}\label{equilibrium-1}
\addcontentsline{toc}{subsubsection}{Equilibrium}

A stationary equilibrium is a tuple \(\{V(b), W, s, M(b)\}\) such that

\begin{enumerate}
\def\labelenumi{\arabic{enumi}.}
\tightlist
\item
  \(W\) and \(V(b)\) satisfy Eq.~\ref{eq-hjb-single} and
  Eq.~\ref{eq-hjb-married} with the same marriage/divorce threshold
  \(b^*\), and
\item
  \(s\) and \(M(b)\) constitute a stationary distribution satisfying
  Eq.~\ref{eq-kf} with Eq.~\ref{eq-population-ct} for all \(b > b^*\).
\end{enumerate}

\section{Estimation}\label{sec-estimation}

\subsection{Calibration}\label{calibration}

For calibration, I use the parameter values in Greenwood and Guner
(\citeproc{ref-greenwood2009}{2009}), summarized in
Table~\ref{tbl-params-gg09}. Some parameters are set from a priori
information: \(\tilde{\beta} = 0.96\) (convention), \(\phi = 0.766\)
(OECD equivalence scale), \(\theta = 0.206\) and \(\kappa = 0.189\)
(\citeproc{ref-mcgrattan1997}{McGrattan et al. 1997}), and
\(1/\delta = 47\) (marriageable life span, ages 18--64). The wage
\(w_{1950}\) is normalized to 1, with a growth rate \(\Delta w = 0.022\)
matching the real wage trend from the data. The remaining ten
parameters, taste parameters \((\alpha, \zeta, \bar{c})\), price
parameters \((p_{1950}, \Delta p)\), and match quality parameters
\((\mu_s, \sigma_s, \mu_m, \sigma_m, \varrho)\), are jointly estimated
by minimum distance, targeting 16 data observations across five
dimensions: time allocations for married and single households
(1950--1990), the fraction married, divorce rate, and marriage rate
(1950 and 2000). Importantly, the match quality parameters do not enter
the labor supply functions, so they are effectively identified from the
vital statistics targets alone.

\begin{table}

\caption{\label{tbl-params-gg09}Parameter Values in Greenwood and Guner
(\citeproc{ref-greenwood2009}{2009})}

\centering{

\centering
\begin{tblr}[         
]                     
{                     
colspec={Q[]Q[]Q[]},
hline{2}={1-3}{solid, black, 0.05em},
hline{1}={1-3}{solid, black, 0.1em},
hline{10}={1-3}{solid, black, 0.1em},
column{1-3}={}{halign=l},
}                     
& Parameter & Description \\
Tastes & $\tilde{\beta}=0.96, \phi=0.766$ & Literature \\
& $\alpha = 0.278, \zeta = -1.901, \overline{c} = 0.131$ & Estimated \\
Technology & $\theta = 0.206, \kappa = 0.189$ & Literature \\
Life span & $1/\delta=47$ & Settings \\
Shocks & $\mu_s = -4.252, \sigma_s^2 = 8.063$ & Estimated \\
& $\mu_m = 0.521, \sigma_m^2 = 0.680, \varrho = 0.896$ & Estimated \\
Prices & $p_{1950} = 9.959, \Delta p = 0.059$ & Estimated \\
Wages & $w_{1950} = 1.000, \Delta w = 0.022$ & Data \\
\end{tblr}

}

\end{table}%

\subsubsection*{Continuous-time Specific
Parameters}\label{continuous-time-specific-parameters}
\addcontentsline{toc}{subsubsection}{Continuous-time Specific
Parameters}

The exit rate \(\nu\) and the discount rate \(\tilde{\rho}\) are
obtained from their discrete-time counterparts via the standard mapping
\(\nu = -\log(1-\delta)/\Delta t\) and
\(\tilde{\rho} = -\log(\tilde{\beta})/\Delta t\), giving the effective
discount rate \(\rho = \tilde{\rho} + \nu\). I set the Poisson arrival
rate \(\lambda = 1\) and the model period length \(\Delta t = 1\) for
the continuous-time model. This choice corresponds exactly to the
discrete-time assumption that each single meets one potential partner
per period, since the expected number of Poisson arrivals in an interval
of length \(\Delta t\) is \(\lambda\,\Delta t = 1\). Moreover, the HJB
equations depend on rates only through their ratios, so the level of
\(\lambda\) is not separately identified; what matters is the product
\(\lambda\,\Delta t\). Setting \(\lambda = 1/\Delta t\) fixes this
product at unity for any \(\Delta t\), making the choice of period
length immaterial.

The continuous-time method requires additional parameters for the OU
process. A naive approach is to set the parameters of the OU process to
match the mean and variance of the AR(1) process. However, this approach
does not work well in practice because of the \emph{continuous
monitoring problem}. In the discrete-time model, the divorce decision is
evaluated only once per period: even if the match quality dips below the
threshold within the period, the couple remains married as long as the
realized end-of-period quality exceeds the threshold. In continuous
time, by contrast, the value function is monitored at every instant, so
the process can trigger divorce the moment it crosses the threshold, an
event that occurs with strictly higher probability. As a result, naively
matching the OU parameters \((\mu_m, \sigma_m, \eta)\) to the AR(1)
parameters \((\mu_m, \sigma_m, \varrho)\) via the standard mapping
\(\eta = -\log(\varrho)/\Delta t\) systematically overstates the divorce
rate in the continuous-time formulation.

To address this issue, I re-estimate the OU parameters
\((\mu_m, \sigma_m, \eta)\) by minimum distance, targeting the fraction
married, divorce rate, and marriage rate in 1950 and 2000. The resulting
parameter values are reported in Table~\ref{tbl-params-ct} alongside the
naive AR(1)-matched values from Greenwood and Guner
(\citeproc{ref-greenwood2009}{2009}), and the model fit is shown in
Table~\ref{tbl-calibration}.

\begin{table}

\caption{\label{tbl-params-ct}Additional Parameters for Continuous-time
Method}

\centering{

\centering
\begin{talltblr}[         
entry=none,label=none,
note{}={Notes: The Greenwood and Guner (2009) column uses the original values of $\mu_m$ and $\sigma_m^2$ and sets $\eta = -\log(\varrho)/\Delta t$ with $\varrho = 0.896$ and $\Delta t = 1$. The Estimated column shows the values re-estimated by minimum distance targeting the fraction married, divorce rate, and marriage rate in 1950 and 2000.},
]                     
{                     
width={0.8\linewidth},
colspec={X[0.15]X[0.35]X[0.3]},
hline{2}={1-3}{solid, black, 0.05em},
hline{1}={1-3}{solid, black, 0.1em},
hline{5}={1-3}{solid, black, 0.1em},
column{1}={}{halign=c},
column{2-3}={}{halign=r},
}                     
Parameter & Greenwood and Guner (2009) & Estimated \\
$\mu_m$ & 0.521 & 0.951 \\
$\sigma_m^2$ & 0.68 & 0.83 \\
$\eta$ & 0.11 & 0.113 \\
\end{talltblr}

}

\end{table}%

The estimated parameters in Table~\ref{tbl-params-ct} are consistent
with the continuous monitoring effect described above. The
mean-reversion speed \(\eta\) hardly changes, but \(\mu_m\) rises from
\(0.521\) to \(0.951\), shifting the match quality distribution away
from the divorce boundary to offset the higher crossing probability
under continuous monitoring.

\subsection{Results}\label{results}

\begin{figure}

\centering{

\includegraphics[width=0.9\linewidth,height=\textheight,keepaspectratio]{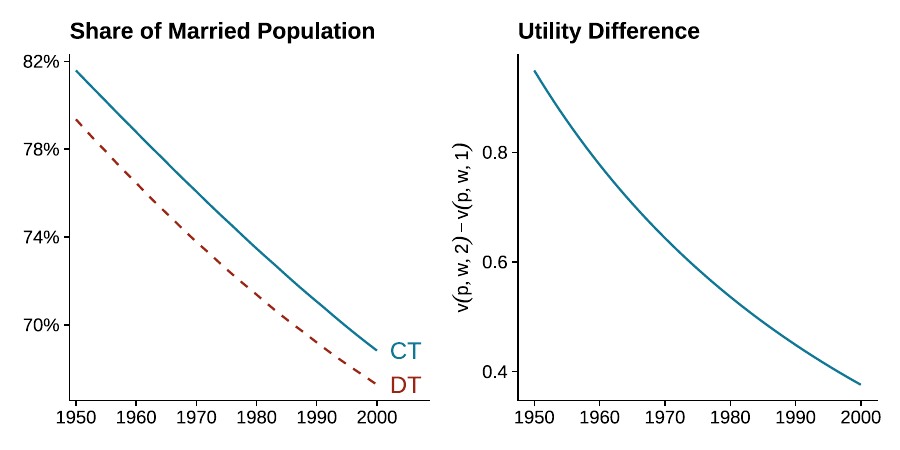}

}

\caption{\label{fig-marriage-rate}\textbf{Replication of Fig. 6 in
Greenwood and Guner (\citeproc{ref-greenwood2009}{2009}).} The left
panel shows the share of married population under the CT and DT methods.
The right panel plots the utility difference \(v(p,w,2) - v(p,w,1)\).}

\end{figure}

Following Greenwood and Guner (\citeproc{ref-greenwood2009}{2009}), I
simulate the model from 1950 to 2020 by increasing the wage \(w\) at
rate \(\Delta w = 0.022\) and decreasing the price of home goods \(p\)
at rate \(\Delta p = 0.059\) (see Table~\ref{tbl-params-gg09}). The
rising wage raises the opportunity cost of home production, while the
falling relative price of home goods makes market substitutes more
accessible. Together, these secular trends compress the utility gain
from marriage \(v(p,w,2) - v(p,w,1)\), as shown in the right panel of
Figure~\ref{fig-marriage-rate}. As the surplus from marriage shrinks,
the equilibrium share of married households declines steadily from
roughly 80 \% in 1950 to below 70 \% by 2020 (left panel of
Figure~\ref{fig-marriage-rate}).

\begin{figure}

\centering{

\includegraphics[width=0.9\linewidth,height=\textheight,keepaspectratio]{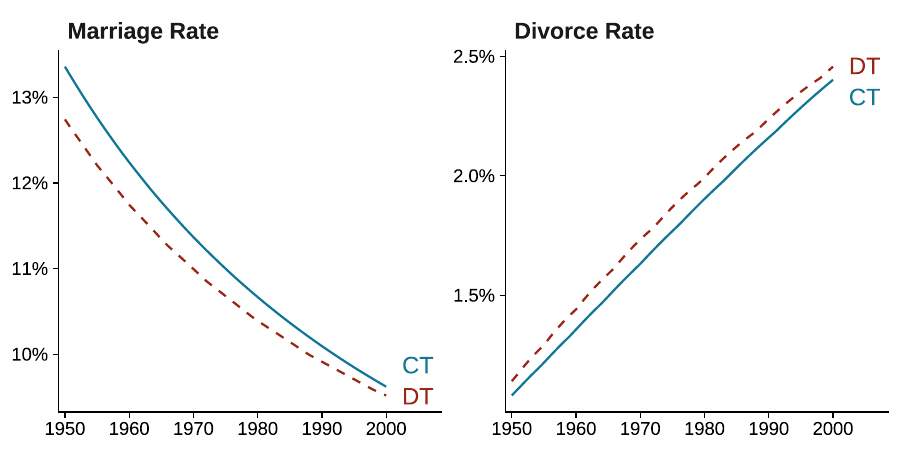}

}

\caption{\label{fig-marriage-divorce-rate}\textbf{Replication of Fig. 7
in Greenwood and Guner (\citeproc{ref-greenwood2009}{2009}).} CT hazard
rates are converted to annual probabilities via \(1 - e^{-r}\) for
comparability with DT.}

\end{figure}

Figure~\ref{fig-marriage-divorce-rate} decomposes the decline in the
married population into its flow components. Since the continuous-time
model yields instantaneous hazard rates, I convert them to annual
probabilities via \(1 - e^{-r}\) (where \(r\) is the hazard rate and
\(\Delta t = 1\) year) so that the CT and DT series are directly
comparable. Both the divorce rate and the marriage rate exhibit the same
broad pattern across the CT and DT formulations: the divorce rate rises
while the marriage rate falls, consistent with the shrinking marriage
surplus. The CT and DT trajectories track each other closely, confirming
that the continuous-time reformulation preserves the aggregate dynamics
of the original model.

\subsubsection*{Computational
Performance}\label{computational-performance}
\addcontentsline{toc}{subsubsection}{Computational Performance}

\begin{figure}

\centering{

\includegraphics[width=0.9\linewidth,height=\textheight,keepaspectratio]{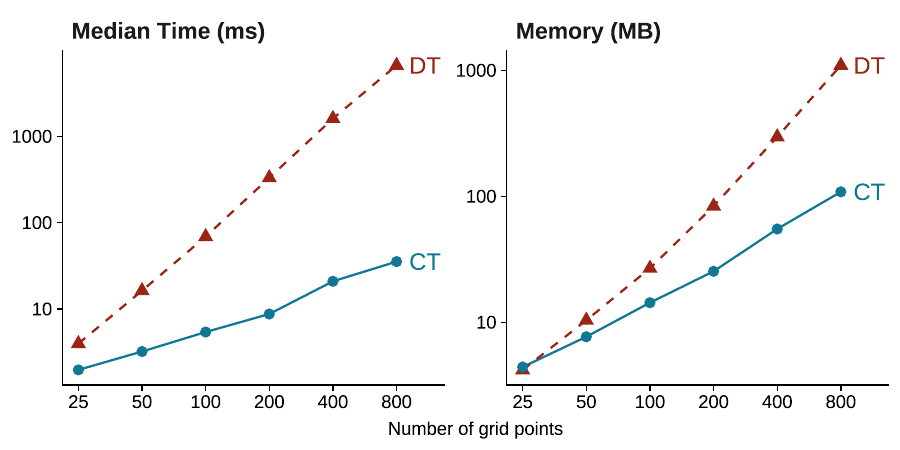}

}

\caption{\label{fig-benchmark}\textbf{Computational Performance of
Continuous-Time and Discrete-Time Methods.} Benchmarked with
BenchmarkTools.jl (\citeproc{ref-chen2016}{Chen and Revels 2016}) on an
Apple M3 Max (16-core, 128 GB RAM).}

\end{figure}

Although the CT and DT methods produce nearly identical equilibrium
outcomes, their computational costs differ markedly. As discussed in
Section~\ref{sec-comp-complexity}, the DT method requires constructing
and iterating on an \(N \times N\) transition matrix, yielding
\(O(N^2)\) complexity in the number of grid points \(N\), whereas the CT
method solves tridiagonal systems that scale as \(O(N)\). The same
distinction applies to memory: the DT method stores the dense
\(N \times N\) Tauchen matrix, requiring \(O(N^2)\) memory, while the CT
method stores only the tridiagonal bands, requiring \(O(N)\). I measure
computation time and memory usage with BenchmarkTools.jl
(\citeproc{ref-chen2016}{Chen and Revels 2016}) on an Apple M3 Max
(16-core, 128 GB RAM). Figure~\ref{fig-benchmark} plots the results on
log--log axes. The DT curves have a slope of approximately 2, and the CT
curves have a slope of approximately 1, consistent with the theoretical
complexity orders.

\section{Conclusion}\label{sec-conclusion}

This paper reformulates the Greenwood and Guner
(\citeproc{ref-greenwood2009}{2009}) marriage and divorce model in
continuous time using the HACT methods of Achdou et al.
(\citeproc{ref-achdou2022}{2022}). The continuous-time formulation
replaces the AR(1) match quality process with an Ornstein--Uhlenbeck
process, the Bellman equations with HJB equations incorporating a
variational inequality for endogenous divorce, and the transition matrix
iteration with a Kolmogorov Forward equation. By exploiting the
tridiagonal structure of the OU generator, all numerical operations
scale as \(O(N)\) in the number of grid points, compared to \(O(N^2)\)
for the discrete-time Tauchen-based approach.

The quantitative exercises show that the continuous-time method closely
tracks the discrete-time equilibrium: the share of married households,
the marriage rate, and the divorce rate all follow nearly identical
paths from 1950 to 2020 under the same secular trends in wages and the
price of home goods. Benchmark comparisons confirm the theoretical
complexity gains, with computation time and memory scaling linearly
rather than quadratically in the grid size.

There are two directions worth highlighting for future work. First, the
current model features exogenous wage and price paths; embedding the
continuous-time marriage block into a general equilibrium framework with
endogenous prices would be a natural extension, and the \(O(N)\) scaling
of the continuous-time solver makes this computationally more tractable.
Second, richer models of marriage and divorce, such as Reynoso
(\citeproc{ref-reynoso2024}{2024}), who incorporates endogenous marriage
and divorce into an equilibrium framework, or models with human capital
accumulation during marriage, add continuous state variables that make
the discrete-time approach increasingly costly. The linear scaling of
the continuous-time solver in each dimension would make these extensions
substantially more feasible.

\newpage{}

\appendix
\renewcommand{\thetable}{\Alph{section}.\arabic{table}}
\setcounter{table}{0}
\renewcommand{\thefigure}{\Alph{section}.\arabic{figure}}
\setcounter{figure}{0}

\begin{center}
{\huge\sffamily\bfseries Appendix}
\end{center}
\vspace{1em}

\section{Computational Details}\label{sec-apdx-comp}

\subsection{Discrete-time Model}\label{discrete-time-model-1}

Computation of the discrete-time model proceeds in three steps: (i)
household allocation, (ii) value function iteration, and (iii)
stationary distributions. Since the household allocation is a static
problem, the indirect utility functions \(v(p, w, 1)\) and
\(v(p, w, 2)\) need only be computed once for each set of prices.

The match quality \(b\) is discretized on an equally spaced grid
\(\{b_1, \dots, b_N\}\) with \(N\) points. The value function \(V(b)\)
evaluated at each grid point is collected into a vector
\(\mathbf{V} = (V_1, \dots, V_N)^{\top}\). The transition kernel of
match quality during marriage, \(G(b'|b)\), is approximated by an
\(N \times N\) row-stochastic matrix \(\mathbf{G}\) using Tauchen
(\citeproc{ref-tauchen1986}{1986})'s method, where
\(G_{j,i} \approx \Pr(b' \in \text{bin } i \mid b = b_j)\). Similarly,
the distribution of match quality for new encounters, \(F(b)\), is
discretized into a probability vector
\(\mathbf{F} = (F_1, \dots, F_N)^{\top}\) with
\(F_i = \Pr(b \in \text{bin } i)\).

\subsubsection*{Value Functions}\label{value-functions-2}
\addcontentsline{toc}{subsubsection}{Value Functions}

The discretized Bellman equations are

\[
\begin{aligned}
\mathbf{V} &= v(p, w, 2) + \mathbf{b} + \beta \, \mathbf{G} \max\{\mathbf{V}, W\},\\
W &= v(p, w, 1) + \beta \, \mathbf{F}^{\top} \max\{\mathbf{V}, W\},
\end{aligned}
\] where \(\mathbf{b} = (b_1, \dots, b_N)^{\top}\) is the grid vector
and \(\max\{\mathbf{V}, W\}\) is taken element-wise. These equations are
solved by standard value function iteration (VFI).

\begin{algorithm}[Value Function
Iteration]\protect\hypertarget{alg-vfi}{}\label{alg-vfi}

~

\begin{enumerate}
\def\labelenumi{\arabic{enumi}.}
\tightlist
\item
  Initialize \(\mathbf{V}\) from a heuristic guess and
  \(W \leftarrow v(p, w, 1)/(1-\beta)\).
\item
  Update \(\mathbf{V}\) and \(W\) simultaneously using the Bellman
  equations above until
  \(\|\mathbf{V}^{n+1} - \mathbf{V}^n\|_\infty < \varepsilon\) and
  \(|W^{n+1} - W^n| < \varepsilon\).
\end{enumerate}

\end{algorithm}

\subsubsection*{Stationary
Distributions}\label{stationary-distributions}
\addcontentsline{toc}{subsubsection}{Stationary Distributions}

The stationary distribution is characterized by an \((N+1)\)-dimensional
state vector, where states \(1, \dots, N\) represent married households
at each grid point and state \(N+1\) represents singles. Let
\(\mathbf{M} = (M_1, \dots, M_N)^{\top}\) denote the mass of married
households at each grid point and \(s\) the mass of singles. With the
exit rate \(\delta\), the stationarity condition is

\[
\begin{pmatrix}
\mathbf{M} \\
s
\end{pmatrix} = (1-\delta)\,\mathbf{P} \begin{pmatrix}\mathbf{M} \\
s
\end{pmatrix} + \begin{pmatrix}\mathbf{0} \\ \delta
\end{pmatrix},
\] where \(\mathbf{P}\) is the \((N+1) \times (N+1)\) column-stochastic
transition matrix defined below. This linear system has the closed-form
solution

\[
\begin{pmatrix}\mathbf{M} \\
s
\end{pmatrix} = \left(\mathbf{I} - (1-\delta)\,\mathbf{P}\right)^{-1} \begin{pmatrix}\mathbf{0} \\ \delta
\end{pmatrix}.
\]

Since \(V(b)\) is increasing in \(b\), there exists a threshold index
\(\iota\) such that

\[
V_{\iota-1} < W < V_{\iota}.
\] Households with match quality above this threshold prefer marriage;
those below prefer being single. To handle the fact that the threshold
generally falls between grid points, I split the bin \(\iota\) using the
interpolation weight

\[
\omega = \frac{W - V_{\iota-1}}{V_{\iota} - V_{\iota-1}} \in [0, 1].
\] A fraction \(\omega\) of bin \(\iota\) is assigned to the
divorce/single state, and the remaining fraction \(1 - \omega\) stays in
the married state. The entries of \(\mathbf{P} = \{P_{i,j}\}\) are

\[
P_{i, j} = \begin{cases}
G_{j, i} & i > \iota, \quad 1 \leq j \leq N \\
G_{j, \iota}\, (1 - \omega) & i = \iota, \quad 1 \leq j \leq N \\
\displaystyle\sum_{k=1}^{\iota-1} G_{j, k} + G_{j, \iota}\, \omega & i = N+1, \quad 1 \leq j \leq N \\[6pt]
F_{i} & i > \iota, \quad j = N+1 \\
F_{\iota}\, (1 - \omega) & i = \iota, \quad j = N+1 \\
\displaystyle\sum_{k=1}^{\iota-1} F_{k} + F_{\iota}\, \omega & i = N+1, \quad j = N+1 \\[6pt]
0 & \text{otherwise.}
\end{cases}
\] Each column sums to one by construction, since
\(\sum_{i=1}^{N} G_{j,i} = 1\) and \(\sum_{i=1}^{N} F_i = 1\).

\subsection{Continuous-time Model}\label{sec-comp-ct}

Computation of the continuous-time model follows Achdou et al.
(\citeproc{ref-achdou2022}{2022}) and proceeds in three steps: (i)
finite difference discretization of the Ornstein--Uhlenbeck (OU)
generator, (ii) solving the HJB equation with a variational inequality
for endogenous divorce, and (iii) solving the Kolmogorov Forward
equation for the stationary distribution.

As in the discrete-time case, the match quality \(b\) is discretized on
a uniform grid \(\{b_1, \dots, b_N\}\) with spacing
\(\Delta b = b_{i} - b_{i-1}\), centered at \(\mu_m\) and spanning
\(\pm n_{\text{std}} \cdot \sigma_m\) standard deviations of the OU
stationary distribution. Value functions evaluated at the grid are
collected into vectors \(\mathbf{V} = (V_1, \dots, V_N)^{\top}\).

\subsubsection*{OU Operator}\label{ou-operator}
\addcontentsline{toc}{subsubsection}{OU Operator}

The OU process Eq.~\ref{eq-ou-process} has drift \(\eta(\mu_m - b)\) and
diffusion coefficient \(\eta\sigma_m^2\). Its infinitesimal generator,
which appears as the differential operator on the right-hand side of the
married HJB Eq.~\ref{eq-hjb-married}, is discretized into a sparse
\(N \times N\) matrix \(\mathbf{A}\) using an upwind scheme for the
drift and central differences for the diffusion. Denoting the drift at
each grid point by \(d_i = \eta(\mu_m - b_i)\), the entries are

\[
A_{i,j} = \begin{cases}
d_i^{+}/\Delta b + \eta\sigma_m^2 / \Delta b^2 & j = i + 1, \\
-d_i^{-}/\Delta b + \eta\sigma_m^2 / \Delta b^2 & j = i - 1, \\
-(A_{i,i-1} + A_{i,i+1}) & j = i, \\
0 & \text{otherwise},
\end{cases}
\] for interior points \(i = 2, \dots, N-1\), where
\(d_i^{+} = \max(d_i, 0)\) and \(d_i^{-} = \min(d_i, 0)\) denote the
positive and negative parts of the drift. The upwind scheme selects the
forward difference when the drift is positive (\(b_i < \mu_m\)) and the
backward difference when it is negative (\(b_i > \mu_m\)), ensuring
monotonicity of the scheme. At the boundaries \(i = 1\) and \(i = N\),
one-sided differences with reflecting conditions are imposed.

The resulting matrix \(\mathbf{A}\) is tridiagonal, so all linear
systems involving \(\mathbf{A}\) can be solved in \(O(N)\) time via
sparse LU factorization.

\subsubsection*{Value Functions}\label{value-functions-3}
\addcontentsline{toc}{subsubsection}{Value Functions}

Evaluating the married HJB Eq.~\ref{eq-hjb-married} at each grid point
and replacing the derivatives with the finite difference operator
\(\mathbf{A}\) gives the system

\[
\rho\,\mathbf{V} = \mathbf{u}_m + \mathbf{A}\,\mathbf{V},
\] where
\(\mathbf{u}_m = (v(p, w, 2) + b_1, \dots, v(p, w, 2) + b_N)^{\top}\) is
the flow utility vector. Rather than inverting
\((\rho\,\mathbf{I} - \mathbf{A})\) directly, an implicit time-stepping
scheme is used to handle the variational inequality \(V(b) \geq W\).
Introducing a pseudo-time step \(\Delta > 0\),

\[
\frac{\mathbf{V}^{n+1} - \mathbf{V}^n}{\Delta} = \mathbf{u}_m + \mathbf{A}\,\mathbf{V}^{n+1} - \rho\,\mathbf{V}^{n+1}.
\] It rearranges to

\[
V^{n+1} = \mathbf{B}^{-1}\underbrace{\left(\mathbf{u}_m + \frac{1}{\Delta}\,\mathbf{V}^n\right)}_{\mathbf{r}^n},
\] where
\(\mathbf{B} = \left(\frac{1}{\Delta} + \rho\right)\mathbf{I} - \mathbf{A}\)
is a tridiagonal matrix. Since \(\mathbf{B}\) does not depend on
\(\mathbf{V}\) or \(W\), its sparse LU factorization is computed once
before the iteration begins.

The single's value \(W\) satisfies Eq.~\ref{eq-hjb-single}, which
rearranges to

\[
W = \frac{v(p, w, 1) + \lambda \int_{b^*}^\infty V(b)\,f(b)\,db}{\rho + \lambda (1 - F(b^*))},
\] where \(F(\cdot)\) and \(f(\cdot)\) are the CDF and PDF of the
singles' draw distribution \(\mathcal{N}(\mu_s, \sigma_s^2)\). To avoid
grid-snapping, I use the \emph{unclamped} solution
\(\widetilde{\mathbf{V}} = \mathbf{B}^{-1}\mathbf{r}\) computed before
the VI projection \(\max\{\cdot, W\}\). Since \(\widetilde{V}(b)\) is
increasing, there exists a threshold index \(\iota\) such that
\(\widetilde{V}_{\iota-1} < W < \widetilde{V}_\iota\), and the
interpolation weight is

\[
\omega = \frac{W - \widetilde{V}_{\iota-1}}{\widetilde{V}_{\iota} - \widetilde{V}_{\iota-1}} \in [0, 1].
\]

The divorce threshold is then \(b^* = b_{\iota-1} + \omega\,\Delta b\).
The acceptance probability is computed as \(1 - F(b^*)\), and the
conditional expectation is approximated by linearly interpolating at the
boundary cell:

\[
\int_{b^*}^\infty V(b)\,f(b)\,db \;\approx\; (1 - \omega)\,V_{\iota-1}\,f(b_{\iota-1})\,\Delta b + \sum_{k=\iota}^{N} V_k\,f(b_k)\,\Delta b.
\]

The algorithm uses a nested loop structure.\footnote{Unlike the DT
  Bellman operator, the CT implicit scheme combined with the variational
  inequality is not a contraction mapping when \(W\) varies
  simultaneously with \(\mathbf{V}\). Fixing \(W\) in the inner loop
  restores monotone convergence of \(\mathbf{V}\). Mellior and Shibayama
  (\citeproc{ref-mellior2024}{2024}) solve related optimal stopping
  problems in HACT models using the splitting method and linear
  complementarity. The marriage model here requires a different
  approach, an inner--outer decomposition, because the single's value
  \(W\) depends endogenously on \(\mathbf{V}\).} The outer loop updates
\(W\) with damping, and the inner loop solves the married HJB to
convergence for a given \(W\):

\begin{algorithm}[HJB iteration with variational
inequality]\protect\hypertarget{alg-hjb}{}\label{alg-hjb}

~

\begin{enumerate}
\def\labelenumi{\arabic{enumi}.}
\tightlist
\item
  Initialize \(W \leftarrow v(p, w, 1)/\rho\) and \(\mathbf{V}\) from a
  heuristic guess.
\item
  \textbf{Inner loop} (holding \(W\) fixed): repeat until
  \(\|\mathbf{V}^{n+1} - \mathbf{V}^n\|_\infty < \varepsilon\):

  \begin{enumerate}
  \def\labelenumii{\alph{enumii}.}
  \tightlist
  \item
    Form
    \(\mathbf{r}^n = \mathbf{u}_m + \frac{1}{\Delta}\,\mathbf{V}^{n}\)
    and compute the unclamped solution
    \(\widetilde{\mathbf{V}}^{n+1} = \mathbf{B}^{-1}\,\mathbf{r}^n\),
    which costs \(O(N)\).
  \item
    Project: \(V_i^{n+1} \leftarrow \max\{\widetilde{V}_i^{n+1},\, W\}\)
    for all \(i\).
  \end{enumerate}
\item
  \textbf{Outer update}: locate \(b^*\) from the unclamped
  \(\widetilde{\mathbf{V}}\) via linear interpolation, update
  \(W^{\text{new}}\), and dampen
  \(W \leftarrow \tfrac{1}{2}W^{\text{new}} + \tfrac{1}{2}W\).
\item
  Repeat steps 2--3 until \(|W^{\text{new}} - W| < \varepsilon\).
\end{enumerate}

\end{algorithm}

\subsubsection*{Stationary
Distribution}\label{stationary-distribution-2}
\addcontentsline{toc}{subsubsection}{Stationary Distribution}

The stationary distribution solves the KFE Eq.~\ref{eq-kf} subject to
the population constraint Eq.~\ref{eq-population-ct}. The threshold
\(b^* = b_{\iota-1} + \omega\,\Delta b\) lies between grid points
\(b_{\iota-1}\) and \(b_\iota\). Rather than solving a single KFE with a
fractionally weighted boundary cell, I solve \emph{two} KFEs at the
adjacent clean grid boundaries and linearly interpolate the resulting
singles fractions.

For a given starting index \(j \in \{\iota - 1, \iota\}\), define the
continuation grid \(\{b_j, b_{j+1}, \dots, b_N\}\) with an absorbing
boundary at \(b_j\). The KFE operator restricted to this grid is the
tridiagonal matrix

\[
\mathbf{T}_j = \mathbf{A}_j^\top - \nu\,\mathbf{I},
\] where \(\mathbf{A}_j\) is the submatrix of \(\mathbf{A}\) on the
continuation grid, and
\(\mathbf{f}_j = \bigl(f(b_j),\, f(b_{j+1}),\, \dots,\, f(b_N)\bigr)^\top\)
is the restricted singles' density. No entries are rescaled; the
boundary falls exactly on a grid point. The KFE and population
constraint combine into

\begin{equation}\protect\phantomsection\label{eq-kf-population}{
\begin{pmatrix}
\mathbf{T}_j & \lambda\,\mathbf{f}_j \\
\Delta b\,\mathbf{1}^\top & 1
\end{pmatrix}
\begin{pmatrix}
\mathbf{M}_j \\
s_j
\end{pmatrix} =
\begin{pmatrix}
\mathbf{0} \\
1
\end{pmatrix},
}\end{equation} which is solved via the Schur complement. Let
\(\mathbf{z}_j = \mathbf{T}_j^{-1}\mathbf{f}_j\) (a single tridiagonal
solve). Then

\[
s_j = \frac{1}{1 - \lambda\,\Delta b\,\sum_i z_{j,i}}.
\]

The two solves yield \(s_{\iota-1}\) (larger continuation region, lower
\(s\)) and \(s_{\iota}\) (smaller continuation region, higher \(s\)).
The interpolated singles fraction is

\[
s = (1 - \omega)s_{\iota-1} + \omega\,s_{\iota}.
\]

Because each solve uses an unmodified tridiagonal system, the solution
varies smoothly in the grid size \(N\) and in \(b^*\). The
fractional-weighting approach of scaling a boundary row by
\((1-\omega)\) introduces a discontinuity whenever the threshold index
\(\iota\) jumps, since the matrix dimension changes discretely; the
two-solve interpolation avoids this artifact. The cost is two \(O(N)\)
tridiagonal solves instead of one, which is negligible compared to the
HJB iteration.

\subsection{Computational Complexity}\label{sec-comp-complexity}

The key difference in computational cost between the two methods stems
from how the match quality transition is represented. In the
discrete-time model, the Tauchen approximation produces a dense
\(N \times N\) row-stochastic matrix \(\mathbf{G}\). Each VFI iteration
requires the matrix-vector product \(\mathbf{G}\max\{\mathbf{V}, W\}\),
which costs \(O(N^2)\). The stationary distribution involves the same
dense matrix through the transition matrix \(\mathbf{P}\), so the
overall complexity of the discrete-time solver is \(O(N^2)\).

In the continuous-time model, the upwind finite difference
discretization of the OU generator produces a tridiagonal matrix
\(\mathbf{A}\), so the implicit matrix
\(\mathbf{B} = \bigl(\frac{1}{\Delta} + \rho\bigr)\mathbf{I} - \mathbf{A}\)
is also tridiagonal. Its sparse LU factorization is computed once and
reused across all HJB iterations, each of which costs \(O(N)\). The KFE
requires two additional \(O(N)\) tridiagonal solves. The overall
complexity of the continuous-time solver is therefore \(O(N)\) in the
grid size.

\section{Supplemental Figures and
Tables}\label{supplemental-figures-and-tables}

\begin{table}

\caption{\label{tbl-calibration}Calibration Targets and Fit}

\centering{

\centering
\begin{talltblr}[         
entry=none,label=none,
note{}={Notes: Data targets are from Greenwood and Guner (2009, Table 3). DT: discrete-time model solved by VFI with the Tauchen transition matrix. CT: continuous-time model solved via the HJB variational inequality and KFE. The CT match quality parameters are re-estimated to match the data targets; all other parameters are shared.},
]                     
{                     
width={0.8\linewidth},
colspec={X[0.2]X[0.1]X[0.1]X[0.1]X[0.1]X[0.1]X[0.1]},
hline{2}={3,6-7}{solid, black, 0.05em},
hline{3}={1-7}{solid, black, 0.05em},
hline{2}={2,5}{solid, black, 0.05em, l=-0.5},
hline{2}={4}{solid, black, 0.05em, r=-0.5},
hline{1}={1-7}{solid, black, 0.1em},
hline{6}={1-7}{solid, black, 0.1em},
column{3-4,6-7}={}{halign=c},
cell{1}{1}={}{halign=c},
cell{1}{2}={c=3}{halign=c, wd=0.3\linewidth},
cell{1}{5}={c=3}{halign=c, wd=0.3\linewidth},
cell{2-5}{1}={}{halign=l},
cell{2-5}{2}={}{halign=c},
cell{2-5}{5}={}{halign=c},
}                     
& 1950 &  &  & 2000 &  &  \\
& Data & CT & DT & Data & CT & DT \\
Frac. married & 0.816 & 0.807 & 0.794 & 0.625 & 0.677 & 0.673 \\
Prob. divorce & 0.011 & 0.012 & 0.011 & 0.023 & 0.026 & 0.025 \\
Prob. marriage & 0.211 & 0.131 & 0.127 & 0.082 & 0.096 & 0.095 \\
\end{talltblr}

}

\end{table}%

\newpage{}

\section*{References}\label{references}
\addcontentsline{toc}{section}{References}

\protect\phantomsection\label{refs}


\begin{CSLReferences}{1}{1}
\bibitem[\citeproctext]{ref-achdou2022}
Achdou, Yves, Jiequn Han, Jean-Michel Lasry, Pierre-Louis Lions, and
Benjamin Moll. 2022. {``Income and {Wealth Distribution} in
{Macroeconomics}: {A Continuous-Time Approach}.''} \emph{The Review of
Economic Studies} 89 (1): 45--86.
\url{https://doi.org/10.1093/restud/rdab002}.

\bibitem[\citeproctext]{ref-chen2016}
Chen, Jiahao, and Jarrett Revels. 2016. \emph{Robust Benchmarking in
Noisy Environments}. arXiv:1608.04295. arXiv.
\url{https://doi.org/10.48550/arXiv.1608.04295}.

\bibitem[\citeproctext]{ref-greenwood2009}
Greenwood, Jeremy, and Nezih Guner. 2009. {``Marriage and {Divorce}
Since {World War II}: {Analyzing} the {Role} of {Technological Progress}
on the {Formation} of {Households}.''} In \emph{{NBER Macroeconomics
Annual} 2008, {Volume} 23}. University of Chicago Press.

\bibitem[\citeproctext]{ref-greenwood2016}
Greenwood, Jeremy, Nezih Guner, Georgi Kocharkov, and Cezar Santos.
2016. {``Technology and the {Changing Family}: {A Unified Model} of
{Marriage}, {Divorce}, {Educational Attainment}, and {Married Female
Labor-Force Participation}.''} \emph{American Economic Journal:
Macroeconomics} 8 (1): 1--41.
\url{https://doi.org/10.1257/mac.20130156}.

\bibitem[\citeproctext]{ref-kaplan2018}
Kaplan, Greg, Benjamin Moll, and Giovanni L. Violante. 2018. {``Monetary
{Policy According} to {HANK}.''} \emph{American Economic Review} 108
(3): 697--743. \url{https://doi.org/10.1257/aer.20160042}.

\bibitem[\citeproctext]{ref-mcgrattan1997}
McGrattan, Ellen R., Richard Rogerson, and Randall Wright. 1997. {``An
{Equilibrium Model} of the {Business Cycle} with {Household Production}
and {Fiscal Policy}.''} \emph{International Economic Review} 38 (2):
267--90. \url{https://doi.org/10.2307/2527375}.

\bibitem[\citeproctext]{ref-mellior2024}
Mellior, Gustavo, and Katsuyuki Shibayama. 2024. {``Solving {HACT}
Models with Bankruptcy Choice.''} \emph{Economics Letters} 245
(December): 112045. \url{https://doi.org/10.1016/j.econlet.2024.112045}.

\bibitem[\citeproctext]{ref-reynoso2024}
Reynoso, Ana. 2024. {``The {Impact} of {Divorce Laws} on the
{Equilibrium} in the {Marriage Market}.''} \emph{Journal of Political
Economy} 132 (12): 4155--204. \url{https://doi.org/10.1086/732532}.

\bibitem[\citeproctext]{ref-tauchen1986}
Tauchen, George. 1986. {``Finite State Markov-Chain Approximations to
Univariate and Vector Autoregressions.''} \emph{Economics Letters} 20
(2): 177--81. \url{https://doi.org/10.1016/0165-1765(86)90168-0}.

\bibitem[\citeproctext]{ref-voena2015}
Voena, Alessandra. 2015. {``Yours, {Mine}, and {Ours}: {Do Divorce Laws
Affect} the {Intertemporal Behavior} of {Married Couples}?''}
\emph{American Economic Review} 105 (8): 2295--332.
\url{https://doi.org/10.1257/aer.20120234}.

\end{CSLReferences}
\end{document}